\documentclass[aps,amsfonts,amsmath,prd,preprint,nofootinbib]{revtex4}
\newcommand{\beq}{\begin{equation}}
\newcommand{\eeq}{\end{equation}}

\usepackage{epsfig,bbm,cancel}
\usepackage{hyperref}

\begin{document}

\title{Higher-dimensional DBI Solitons}

\author{Handhika S. Ramadhan}
\email{handhika@cosmos.phy.tufts.edu, ramadhan@teori.fisika.lipi.go.id}
\affiliation{Institute of Cosmology, Dept.~of Physics and Astronomy Tufts University\\ Medford, MA 02155, USA\\ and\\ Group for Theoretical and Computational Physics\footnote{Present affiliation.}, Indonesian Institute of Sciences, Kompleks Puspiptek Serpong, Tangerang 15310, Indonesia}

\def\changenote#1{\footnote{\bf #1}}

\begin{abstract}
We study the theory of a (global) texture with DBI-like Lagrangian, the higher-dimensional generalization of the previously known {\it chiral Born-Infeld theory}. This model evades Derrick's theorem and enables the existence of solitonic solutions in arbitrary $(N+1)$-dimensions. We explicitly show the solutions in spherically-symmetric ansatz. These are examples of {\it extended} topological solitons. We then investigate the coupling of this theory to gravity, and obtain the static self-gravitating solitonic $p$-brane solutions. These non-singular branes can be identified as the smooth versions of cosmic $p$-branes which, in the thin-wall limit, suffers from naked singularities.
\end{abstract}

\maketitle
\thispagestyle{empty}
\section{Introduction}
\setcounter{page}{1}

Topological defects are a subclass of solitons which are static and owe their existence to their non-trivial boundary conditions (topology). Examples of defects are {\it domain walls, cosmic strings, monopoles}, and {\it textures}~\cite{vilenkinshellard}. They are classified based on the homotopy group of the corresponding vacuum manifold, $\pi_{N}(\mathcal{M})$. In cosmology, topological defects attract much attention since they can naturally form due to phase transitions and may have played important roles in the early universe~\cite{kibble, vilenkin1985}.

Textures are defects that arise when the {\it third} homotopy group is non-trivial~\cite{davis, turok}, $\pi_{3}(\mathcal{M})\neq\mathbb{I}$. Often called non-singular solitons (due to the phase of the scalar field which is everywhere well-defined) they are probably best studied in terms of {\it non-linear sigma model}~\cite{misner}. They have been extensively studied in the past due to the possibility of seeding galaxy formation~\cite{davis, turok, turokspergel, stpr}. Derrick's theorem~\cite{derrick}, however, provides an obstacle for the existence of stable textures. In $(3+1)$ dimensions textures are unstable to collapse. The reason is that under spatial rescaling the only minimum static energy configuration is the trivial vacuum and thus nothing prevents them from shrinking. One way to evade this no-go theorem is by having higher order kinetic term(s)\footnote{Recently there has also been several studies on topological defects with non-canonical kinetic term~\cite{noncanon1, noncanon2, noncanon3, noncanon4, noncanon5}.}, as has been shown by Skyrme~\cite{skyrme}, who, by adding a kinetic term $4th$ order in the derivative (called the {\it Skyrme term}), obtained static solitonic solutions dubbed as {\it Skyrmions}. This is possible since the term scales differently than the ordinary kinetic term and as a result the defects have a natural scale. Skyrme interpreted these Skyrmions as baryons with baryon number coming from the topological winding number.

As the number of spatial dimensions increases Derrick's theorem gives more stringent constraint for the defects' stability. In the context of textures the Skyrme term cannot support solitonic solutions in higher-than $(3+1)$ dimensions. In order to have stable {\it extended} textures we have to have generalized higher-order $(>4)$  Skyrme kinetic terms. While it is acceptable as an effective field theory the addition of arbitrary non-canonical terms does not look aesthetic as they unavoidably introduce new scale(s) in the theory which results in new parameters that must be set by hand. A theory with many adjustably-unrelated parameters will lose its attractiveness. Moreover, there is no motivation from fundamental physics of why they should exist. They are simply put by hand to fulfill the purpose of having stable defects. On the other hand, the Skyrme term itself is ad hoc. There is no underlying fundamental principle dictating such a term should exist, other than to have stable solutions. This led people to find generalizations of the Skyrme term that can generate any possible higher order terms~\cite{fujii1, fujii2, marleau1, dion}. One attractive proposal is the so-called {\it chiral Born-Infeld theory}~\cite{chiraldbi}. In this theory, the texture appears in the Born-Infeld-type Lagrangian~\cite{borninfeld} and the higher-order terms are generated automatically via Taylor expansion for small mass-scale parameter. 

In recent years field theories with DBI (Dirac-Born-Infeld) kinetic terms have been intensively studied. They gain interest due to the fact that the appearance of DBI kinetic form can be motivated from fundamental theory ({\it i.e.}, the $D$-brane Action). They appear in the string-theory-inspired inflationary model~\cite{alishahiha} where the inflaton has DBI kinetic term. In topological defects there have been proposals of DBI global strings~\cite{sarangi} or (gauge) cosmic strings~\cite{babichev}. Gravitational field of DBI global monopoles has also been investigated in~\cite{xinzhouli}. All of the discussions focus on the existence of defects in $(3+1)$ dimensions. On the other hand, global $k$-defects in arbitrary $(N+1)$ dimensions have been shown to exist~\cite{avelino}. This gives an evidence of the existence of stable-static defects with non-canonical kinetic terms in higher dimensions. It will particularly be interesting to obtain these {\it extended} defects in DBI field theories. 

From a completely different point of view, in the last decade theories with extra dimensions have extensively been studied in the context of {\it braneworld}. The so-called {\it braneworld scenario} suggests that our $(3+1)$ dimensional universe is a $3$-brane living in a higher-dimensional bulk spacetime~\cite{akama, rubakov, randallsundrum}. Vacuum uncharged $p$-branes solutions in arbitrary $D$ spacetime have been found by Gregroy~\cite{gregory}. These solutions  are boost-symmetric along the branes and asymptotically-flat far from the cores. They, however, suffer from naked singularities. Since phenomenologically the branes can be modeled by topological defects~\cite{cohenkaplan, gregorycompac, ghergetta, olasagasti}, Gregory argued that an appropriate choice of defects core might smooth out the singularity and suggested Skyrme model as an example. Indeed, it has recently been shown that Einstein-Skyrme model in $7d$ has self-gravitating solitonic $3$-branes solutions in its spectrum~\cite{ramadhan}. The solutions are regular at the core and and the energy-momentum tensor falls fast enough outside so that asymptotically approaches the flat vacuum $p$-brane metric. This is an example of codimension-$3$ solitonic defects in higher-dimensional theories. To the best of our knowledge, so far there has not been any discussion of codimension-higher-than-$3$ solitonic defects that can cure the singularities of the thin-wall $p$-branes. This is simply because there are no solitonic defects (with ordinary kinetic term) in codimension higher than three. We might think of using global defects, as in~\cite{olasagasti}, but since their static energy is divergent in flat space their coupling to gravity create deficit angle in the metric and thus cannot be asymptotically flat~\cite{chovilenkin}. 

In this paper we attempt to answer the following two questions. First, can solitonic defects with codimension $n>3$, which we dub {\it extended}-solitons, exist? Second, can we smooth out the singular uncharged $p$-branes in arbitrary dimensions? Using the generalization of chiral Born-Infeld theory~\cite{chiraldbi} we found that the answer for both questions is positive. 

\section{Review of chiral Born-Infeld Theory}

Here we will review the theory of chiral Born-Infeld~\cite{chiraldbi}. As in the Skyrme model~\cite{skyrme} the chiral Born-Infeld theory can be conveniently formulated in terms of a {\it chiral} field $U(t,x)$, a unitary $2\times 2$ scalar matrix transforming under $SU(2)$. The Lagrangian takes the following form\footnote{Here we ignore the $f^{2}_{\pi}$ prefactor as it is only meaningful in the context of effective meson field theory.}
\begin{equation}
\label{eq:chiraldbi}
\mathcal{L}=-\beta^{2}\left(1-\sqrt{1-\frac{1}{2\beta^{2}}Tr\left(L_{\mu}L^{\mu}\right)}\right),
\end{equation}
where $L_{\mu}\equiv U^{\dag}\partial_{\mu}U$ is the {\it left-chiral current} and $\beta$ is a mass dimensional scale parameter of the model. The non-trivial topology manifests itself in the existence of the topological charge~\cite{skyrme}
\begin{equation}
B=\frac{1}{24\pi^{2}}\int d^{3}\epsilon_{ijk}Tr\left(L_{i}L_{j}L_{k}\right).
\end{equation}
At $\beta\rightarrow\infty$ the theory reduces to
\begin{equation}
\label{eq:chiraldbiexpansion}
\mathcal{L}\sim-\frac{1}{4}Tr\left(L_{\mu}L^{\mu}\right),
\end{equation} 
the ordinary {\it non-linear sigma model}. The advantage of this model is that it does not even require the Skyrme term to stabilize the defect. The DBI-form is sufficient to render the texture defects stable.

Considering the spherically-symmetric ansatz
\begin{equation}
U=e^{iF(r)\hat{r}\cdot\bar{\tau}},
\end{equation}
where $F(r)$, the so-called {\it chiral angle}, is a function of radial only and $\bar{\tau}$ are the Pauli matrices, the static energy functional is given by
\begin{equation}
E[F]=8\pi\beta^{2}\int^{\infty}_{0}\left(1-R\right)r^{2}dr,
\end{equation}
with
\begin{equation}
R\equiv\sqrt{1-\frac{1}{\beta^{2}}\left(\frac{F'^{2}}{2}+\frac{\sin^{2}F}{r^{2}}\right)},
\end{equation}
and "{\it primes}" denotes derivative with respect to $r$. The variational principle yields the equation of motion
\begin{equation}
\left(r^{2}\frac{F'}{R}\right)'=\frac{\sin 2F}{R}.  
\end{equation}
The DBI form of the kinetic term ensures that the defects are stable against spatial rescaling and using numerical technique the author of~\cite{chiraldbi} explicitly shows the solitonic solutions for topological charge $B=1$.

\section{Extended Chiral Born-Infeld Solitons}

 \subsection{The Model}
 
From the previous discussion we can learn that since the Lagrangian~(\ref{eq:chiraldbi}) can be expanded to any order there is hope to stabilize the solitonic solutions in any arbitrary dimension. We therefore wish to obtain extended solitons in $(N+1)$-dimensions. For our purpose, it is more convenient to re-write the chiral DBI Lagrangian in terms of {\it non-linear sigma model},
\begin{equation}
\label{eq:chiraldbisigma}
{\cal{L}}_{DBI}=\beta^{2}_{N}\left[\sqrt{1+\frac{1}{\beta^{2}_{N}}\partial_{\mu}\Phi^{i}\partial^{\mu}\Phi^{i}}-1\right],
\end{equation}
where $i=1,2,...,N+1$. Notice that we express the mass scale parameter as $\beta_{N}$, since it is a dimensionful parameter with dimension $\beta_{N}^{2}\sim[M]^{(N+1)}$. It is more convenient to make a change of variables to the dimensionless units,
\begin{equation}
x\rightarrow\frac{x}{M},\  \ \phi\rightarrow M^{\frac{N-1}{2}}\phi,\ \ \beta\rightarrow M^{\frac{N+1}{2}}\beta_{N},
\end{equation}
where $M$ is the mass scale of the theory.

For large $\beta$, the Lagrangian reduces to
\begin{equation}
{\cal{L}}\sim\frac{1}{2}\partial_{\mu}\Phi^{i}\partial^{\mu}\Phi^{i},
\end{equation}
the ordinary non-linear sigma model.

The non-linear constraint $\Phi^{i}\Phi^{i}=1$ spontaneously breaks the symmetry of the theory, $SO(N+1)\rightarrow SO(N)$.  From homotopy theory we know that it defines the vacuum manifold $\mathcal{M}=SO(N+1)/SO(N)\cong S^{N}$, whose $N$-th homotopy group is non-trivial, $\pi_{N}(S^{N})=\mathbb{Z}$. This topological invariant number manifests itself in the topological current
\begin{equation}
\sqrt{-g}j^{\mu}=\frac{1}{12\pi^{2}}\epsilon^{\mu\alpha_{1}\alpha_{2}\ldots\alpha_{N-1}}\epsilon_{a_{1}a_{2}\ldots a_{N}}
\Phi^{a_{1}}\partial_{\alpha_{1}}\Phi^{a_{2}}\partial_{\alpha_{2}}\Phi^{a_{3}}\ldots\partial_{\alpha_{N-1}}\Phi^{a_{N}}.
\end{equation}

 \subsection{Scaling Argument}
 
Having examined the non-trivial topology of the vacuum manifold, we are curious whether there exists a static defect in its spectrum. Let us consider the static energy
\begin{equation}
\label{eq:energy}
E=\int\left(1-\sqrt{1-E_{2}}\right)d^{N}x,
\end{equation}
where, for simplicity, we set $\beta=1$, $E_{2}\equiv\partial_{i}\phi_{a}\partial^{i}\phi_{a}$ is the gradient energy density\footnote{$0<E_{2}<1$ to give a finite and real energy.}. Under the scale transformations, the energy rescales as
\begin{equation}
\label{eq:rescaledenergy}
E_{\lambda}=\int\lambda^{-N}\left(1-\sqrt{1-\lambda^{2}E_{2}}\right)d^{N}x.
\end{equation}
Stability implies~\cite{derrick}
\begin{eqnarray}
\frac{dE}{d\lambda}\bigg|_{\lambda=1}&=&0,\nonumber\\
\frac{d^{2}E}{d\lambda^{2}}\bigg|_{\lambda=1}&>& 0,
\end{eqnarray}
which translates into
\begin{eqnarray}
\label{eq:rescaledconditions}
\frac{E_{2}}{\sqrt{1-E_{2}}}-N(1-\sqrt{1-E_{2}})=0,\nonumber\\
\frac{E_{2}}{(1-E_{2})^{3/2}}+\frac{(1-2N)E_{2}}{\sqrt{1-E_{2}}}+N(N+1)(1-\sqrt{1-E_{2}})\geq0,
\end{eqnarray}
respectively. It is immediately clear that there is no stable defect for\footnote{The $N=2$ case remains conformally-invariant, as in the ordinary non-linear sigma model case, which we do not investigate here.} $N<3$. On the other hand, with some algebra one can prove that for {\it each} $N\geq 3$ there always exists a non-zero positive value of $E_{2}$ that satisfies condition (\ref{eq:rescaledconditions}); {\it i.e.}, the possibility of having static and stable defects are not ruled out in any higher dimensions. 

 \subsection{Hedgehog Ansatz}
 
Armed with the existence of stable defects we consider a spherically-symmetric ({\it hedgehog}) ansatz\footnote{In this paper we only consider $B=1$ topological charge.}, {\it i.e.}, 
\begin{equation}
\Phi_{i}=(\cos\tilde{\theta_{1}},\sin\tilde{\theta_{1}}\cos\tilde{\theta_{2}},
\sin\tilde{\theta_{1}}\sin\tilde{\theta_{2}}\cos\tilde{\theta_{3}},\ldots, \sin\tilde{\theta_{1}}\sin\tilde{\theta_{2}}\sin\tilde{\theta_{3}}\cdots\sin\tilde{\theta_{N}}),
\end{equation}
with 
\begin{eqnarray}
\tilde{\theta_{1}}&=&\alpha(r),\nonumber\\
\tilde{\theta_{j}}&=&\theta_{j}, \ \ \ \ j=2,...,N-1,
\end{eqnarray}  
where $\alpha(r)$ is the {\it chiral angle} and $\theta_{j}$ are the {\it angular} coordinates of an $N$-sphere. The Lagrangian becomes
\begin{equation}
\label{eq:lagchiral}
{\cal{L}}=\beta^{2}\left[\sqrt{1-K}-1\right],
\end{equation}
where
\begin{equation}
\label{eq:K}
K\equiv\frac{1}{\beta^{2}}\left(\alpha'^{2}+\frac{(N-1)\sin^{2}\alpha}{r^{2}}\right),
\end{equation} 
with ``primes" denotes derivative with respect to $r$. This gives the $(N+1)$-dimensional Action
 \begin{eqnarray}
\label{eq:dbiaction}
{\cal{S}}&=&\int d^{N+1}X \ {\cal{L}},\nonumber\\
&=&\Omega_{N-1}\int dt\ dr\ r^{N-1}\ {\cal{L}},
\end{eqnarray}
where $\Omega_{N-1}$ is the hypersurface area of an $(N-1)$-sphere. The Least Action Principle yields the equation for the chiral angle $\alpha(r)$,
\begin{equation}
\label{eq:dbiskyrme}
\left[\frac{\alpha' r^{N-1}}{\sqrt{1-K}}\right]'
-\frac{(N-1)\sin 2\alpha}{2r^{3-N}\sqrt{1-K}}=0.
\end{equation}
It can readily be checked that at weak-coupling limit, $\beta\rightarrow\infty$, the equation reduces to the ordinary (static) texture equation
\begin{equation}
\alpha''+\frac{2\alpha'}{r}-\frac{\sin\alpha}{r^{2}}\approx 0.
\end{equation}

 \subsection{Solitonic DBI Textures in Arbitrary $N$ dimensions}

To obtain unique solutions, eq.(\ref{eq:dbiskyrme}) should be supplemented with appropriate boundary conditions. We impose the following conditions
\begin{eqnarray}
\label{bound}
\alpha(0)=\pi,\nonumber\\
\alpha(\infty)\rightarrow 0.
\end{eqnarray} 
They are chosen to ensure topological charge $B=1$ and finiteness of the energy.

Eq.(\ref{eq:dbiskyrme}) cannot be solved in a closed form. We need to employ numerical method to obtain the solutions. In this paper we use the shooting method to solve it. To have solitonic solutions, we require regularity around the cores. Thus the expansion about $r=0$ is expected to take the following form
\begin{equation}
\alpha(r)=\pi+\alpha_{1}\ r+O(r^{3}),
\end{equation}
where the undetermined coefficient $\alpha_{1}$ becomes the {\it shooting parameter}, {\it i.e.}, the adjustable initial gradient of the function such that the asymptotic solutions satisfy the boundary condition~\ref{bound}. 
\begin{figure}[htbp]
\centering\leavevmode
\epsfysize=9cm \epsfbox{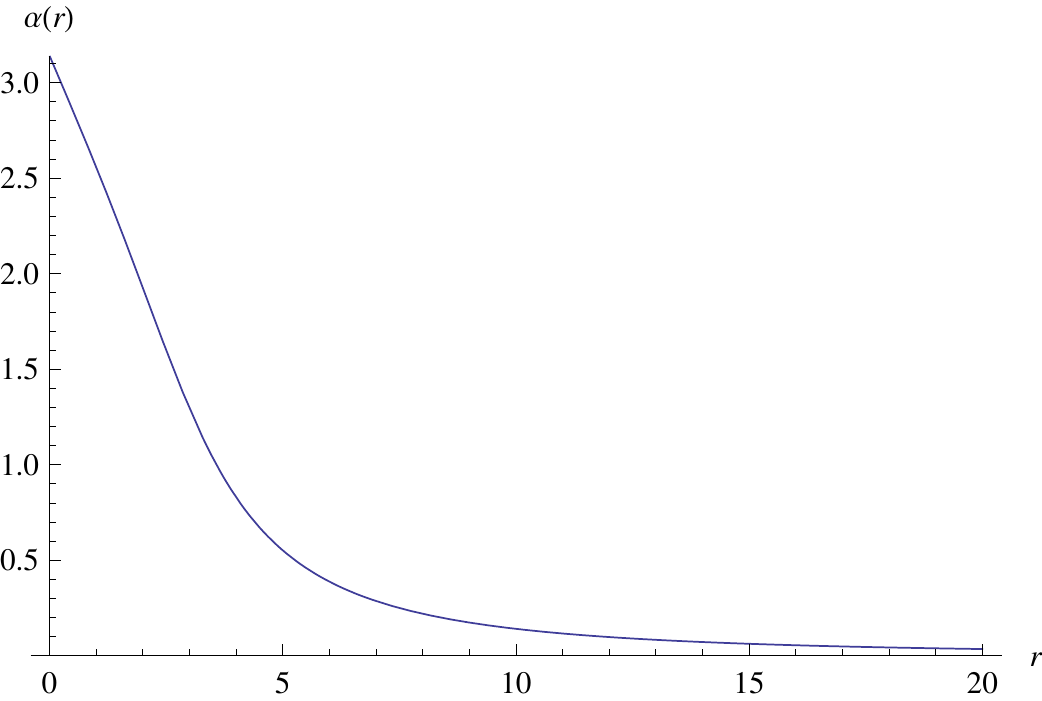}
\caption {Chiral DBI soliton (with $\beta=1$) in flat $N=3$. Here we reproduce the result of~\cite{chiraldbi}.}
\label{gambar}
\end{figure}

\begin{figure}[htbp]
\centering\leavevmode
\epsfysize=10cm \epsfbox{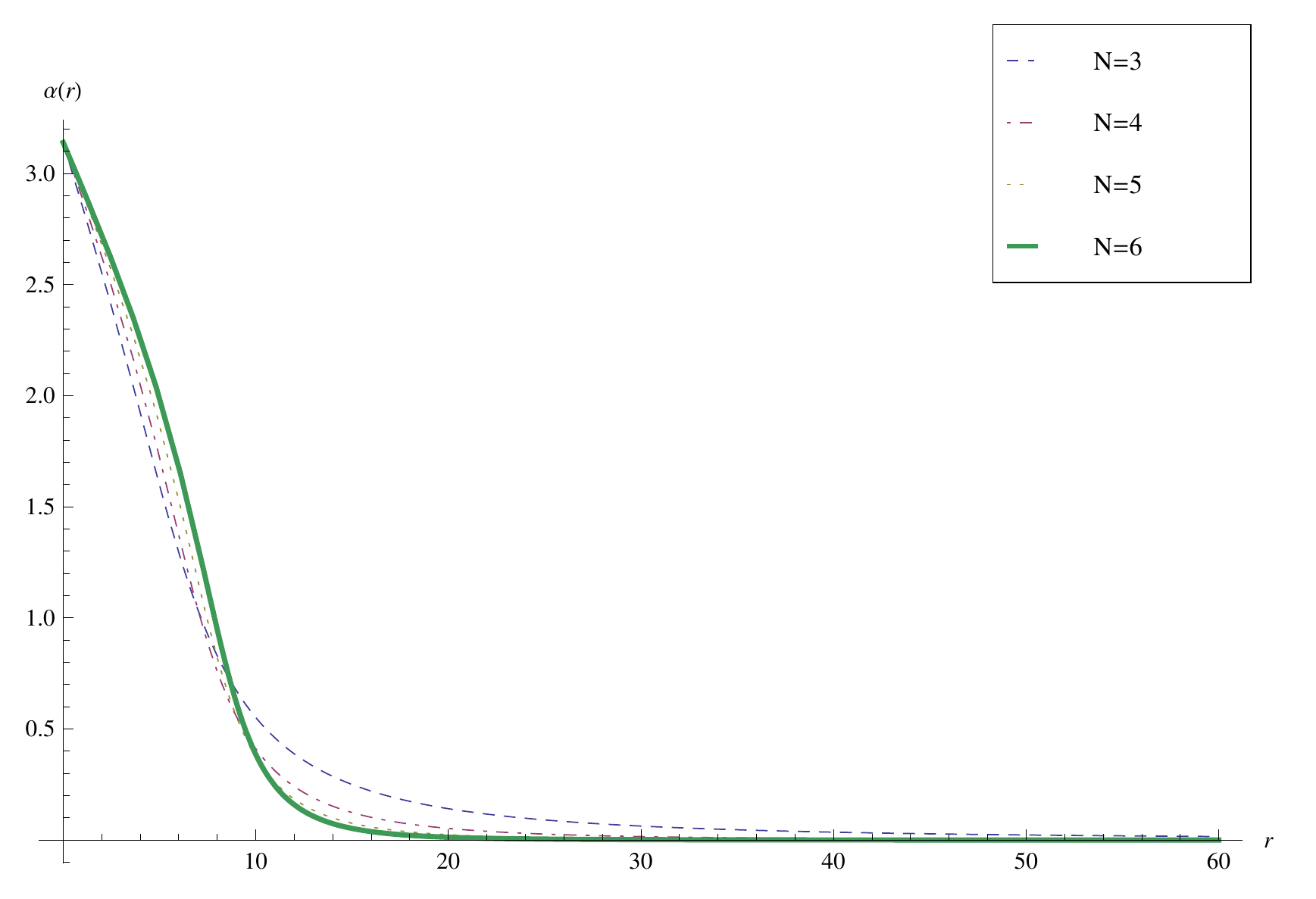}
\caption {Extended DBI solitons (with $\beta=1/2$) in various dimensions ($N=3, 4, 5$, and $6$). The value of $\beta$ is chosen to emphasize the non-linearity of the theory, where $0<\beta<1$.}
\label{chiraldim}
\end{figure}

We obtain solutions for several number of spatial dimensions $N$ and the coupling constant $\beta$. They are the {\it extended} solitons in $(N+1)$-dimensional chiral DBI theory. In fig.\ref{gambar} we reproduce the chiral DBI soliton obtained in~\cite{chiraldbi}. Fig.\ref{chiraldim} shows solutions for various spatial dimensions (up to $N=6$). The profiles are all localized around the cores and quickly relax to zero outside.

From fig.\ref{chiraldim} we can see that the higher-dimensional defect profiles tend to fall-off faster asymptotically than their lower-dimensional counterparts. This fall-off behavior can easily be understood by studying the asymptotic expansion of eq.(\ref{eq:dbiskyrme}). At large radius, the field  can be approximated by
\begin{eqnarray}
\alpha(r)&\approx&\alpha_{\infty}+\delta\alpha(r),\nonumber\\
&\approx&\delta\alpha(r),
\end{eqnarray}
where $|\delta\alpha|\ll 1$ and on the second line we set $\alpha_{\infty}=0$ due to the boundary condition (\ref{bound}). This yields the asymptotic equation
\begin{equation}
\delta\alpha''+(N-1)\frac{\delta\alpha'}{r}\approx 0,
\end{equation}
with (non-trivial) solution
\begin{equation}
\alpha(r)\approx\delta\alpha(r)\propto\frac{b}{r^{N-1}},
\end{equation}  
where $b$ is some numerical constant. As the dimensions increase the inverse power of the polynomial increases, and as a result the chiral angle profile falls off faster.

For any given number of spatial dimensions, the solution is characterized by the coupling constant $\beta$. It is responsible for the size of the defects. We can make some rough estimation about the defect core thickness. The defect thickness is the size that minimizes its static energy configuration. In domain walls, for example, the size can be estimated by the balance of the gradient and potential energy. In our theory, the thickness of defect core is determined by the balance of linear gradient energy term and its non-linear counterparts. The energy is roughly given by
\begin{equation}
E\sim\ell^{N} \beta^{2}\left(\sqrt{1-\frac{1}{\beta^{2}\ \ell^{2}}}-1\right),
\end{equation}
and so the size can roughly be estimated as
\begin{equation}
\ell\sim\frac{(N-1)}{\sqrt{N(N-2)}}\ \beta^{-1}.
\end{equation}

\begin{figure}[htbp]
\centering\leavevmode
\epsfysize=10cm \epsfbox{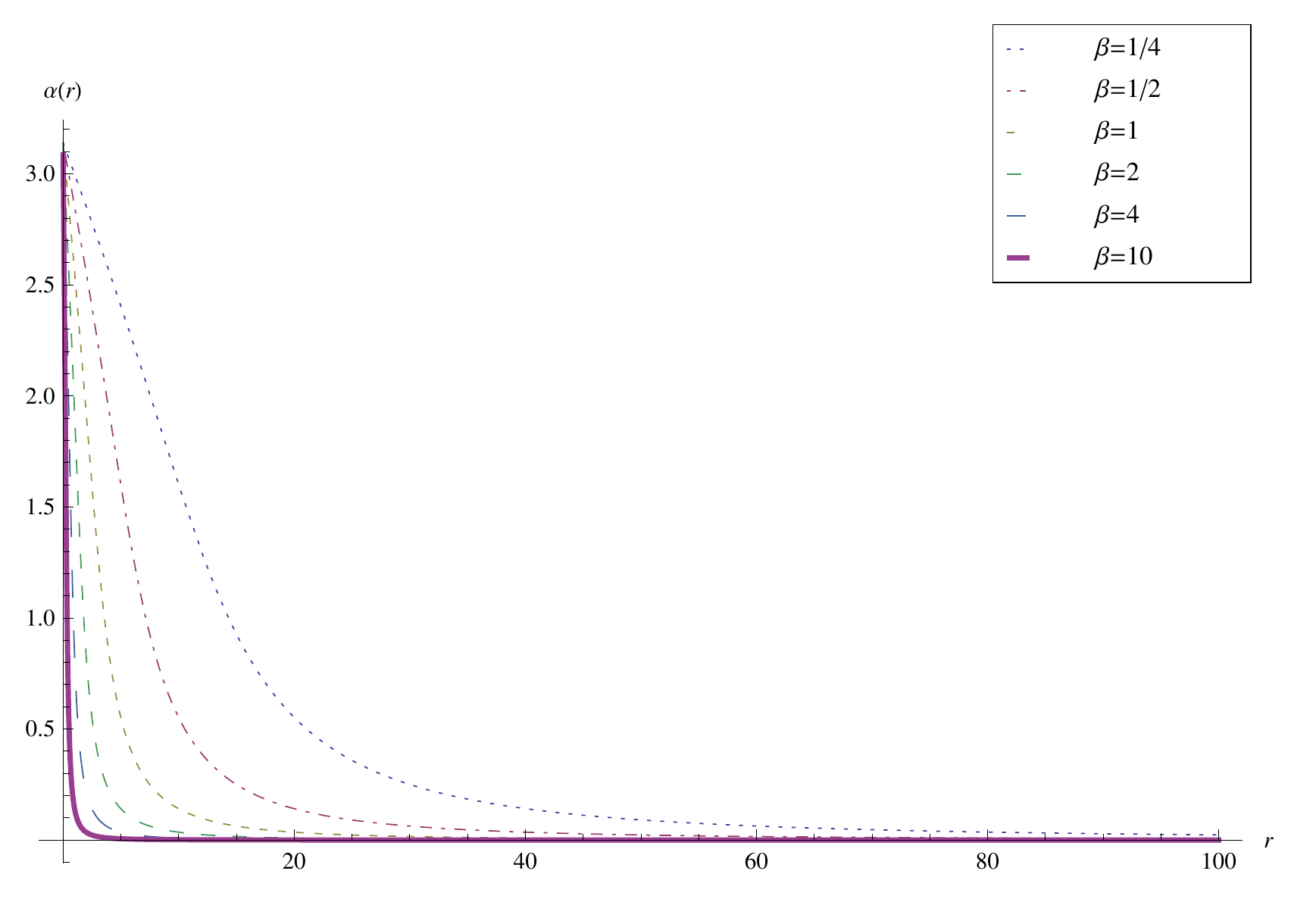}
\caption {$3+1$ dimensional DBI solitons for various values of $\beta$.}
\label{chiralbeta}
\end{figure}

In fig.\ref{chiralbeta} we show solutions for various $\beta$ (in $N=3$). For $\beta>1$ the theory is weakly-coupled and the Lagrangian (\ref{eq:chiraldbisigma}) can be Taylor-expanded to yield the ordinary non-linear sigma model with higher-order kinetic correction terms. On the other hand, for $0<\beta\leq 1$ we can no longer neglect the higher-order expansion terms. The theory becomes strongly-coupled.
One can see from fig.\ref{chiralbeta} that the defect size is inversely proportional to $\beta$; as $\beta$ grows, the defects become thinner. We can understand it from the fact that for large $\beta$ it effectively reduces to an ordinary texture defect, and texture does not possess a natural scale; it is unstable against collapse to trivial vacuum configuration. As $\beta\rightarrow\infty$ the defect is infinitely thin (vacuum).

\section{DBI Texture Branes}

Having established the solitonic solutions in higher-dimensional flat case, we would like to know whether they can be good models for non-singular $p$-brane cores. In other words, we attempt to prove Gregory's conjecture in ~\cite{gregory} that the naked singularity in $p$-branes vacuum solutions can be smoothed-out by an appropriate choice of cores, and the natural candidates would be uncharged topological defects with asymptotically-flat space-time metric. Those conditions put very stringent constraints on the choice of defects, and we will show that the extended DBI texture satisfies all the requirements.

 \subsection{Cosmic $p$-Branes in Isotropic Gauge}
 
Consider $D$-dimensional\footnote{$D=N+p+1$.} $p$-branes in {\it isotropic gauge},
\begin{equation}
\label{eq:isondimensions}
ds^{2}=B(r)^{2}\eta_{\mu\nu}dx^{\mu}dx^{\nu}-H(r)^{2}(dr^{2}+r^{2}d\Omega^{2}_{N-1}),
\end{equation}
where
\begin{equation}
d\Omega^{2}_{N-1}\equiv\sum^{N-1}_{i=1}\Upsilon_{i}(\theta_{j<i})d\theta^{2}_{i},
\end{equation}
with $\Upsilon_{1}\equiv1$ and
\begin{equation}
\Upsilon_{i>1}\equiv\sin^{2}\theta_{1}\prod^{i-1}_{j=2}\sin^{2}\theta_{j},
\end{equation}
and $N$ is the number of the codimensions of the brane. 

The Ricci tensor components are given by
\begin{eqnarray}
\label{eq:einsteiniso}
R^{0}_{0}&=&\left[\frac{B''}{B}+p\left(\frac{B'}{B}\right)^2+(N-2)\frac{B'H'}{BH}+(N-1)\frac{B'}{Br}\right],\nonumber\\
R^{r}_{r}&=&\Bigg[(p+1)\frac{B''}{B}+(N-1)\frac{H''}{H}+(N-1)\frac{H'}{Hr}-(p+1)\frac{B'H'}{BH}\nonumber\\
&&-(N-1)\left(\frac{H'}{H}\right)^2\Bigg],\nonumber\\
R^{\theta}_{\theta}&=&\left[\frac{H''}{H}+(N-3)\left(\frac{H'}{H}\right)^2+(2N-3)\frac{H'}{Hr}+(p+1)\frac{B'H'}{BH}
+(p+1)\frac{B'}{Br}\right].\nonumber\\
\end{eqnarray} 

The vacuum Einstein equation for zeroth component, $R^{0}_{0}=0$, can be written as
\begin{equation}
\left[(B^{p+1})'H^{N-2}r^{N-1}\right]'=0.
\end{equation}
By appropriately identifying the constant of integration of equation above, we could try an ansatz for $B(r)$, generalizing our isotropic solution for $p=N=3$~\cite{ramadhan},:
\begin{equation}
\label{eq:bform}
B(r)^{p+1}=\left(\frac{1-(\frac{r_{o}}{r})^{N-2}}{1+(\frac{r_{o}}{r})^{N-2}}\right)^{m},
\end{equation}
with $r_{o}$ integration constant. This yields
\begin{equation}
\label{eq:hform}
H(r)=\left(\frac{1+(\frac{r_{o}}{r})^{N-2}}{1-(\frac{r_{o}}{r})^{N-2}}\right)^{\frac{m}{N-2}}
\left[1-\left(\frac{r_{o}}{r}\right)^{2(N-2)}\right]^{\frac{1}{N-2}},
\end{equation}
which immediately solves $R^{\theta}_{\theta}=0$. The only left parameter to determine, $m$, can be expressed in terms of variables $p$ and $N$ by plugging (\ref{eq:bform}) and (\ref{eq:hform}) into the remaining equation, $R^{r}_{r}=0$, which then, after some algebra, gives
\begin{equation}
\label{eq:mform}
m=\sqrt{\frac{(p+1)(N-1)}{(N+p-1)}}.
\end{equation}
Eqs (\ref{eq:bform}) and (\ref{eq:hform}), supplemented with (\ref{eq:mform}), then completely solve the vacuum Einstein equations in isotropic form (see also~\cite{corradini}).

 \subsection{The Defect Cores}
 
The full Action for our model is given by
\begin{equation}
{\cal S}=\int d^{D}x\sqrt{-g}\bigg(\frac{R}{2\kappa^{2}}+{\cal L}_{DBI}\bigg).
\end{equation}
The matter-sector Lagrangian still takes the form of (\ref{eq:lagchiral}), with $K$ now redefined as
\begin{equation}
\label{eq:K}
K\equiv\frac{1}{\beta^{2}}\left(\frac{\alpha'^{2}}{H^{2}}+\frac{(N-1)\sin^{2}\alpha}{H^{2}r^{2}}\right).
\end{equation}
The equation for the chiral angle now becomes
\begin{equation}
\label{eq:chiralgrav}
\left[\frac{B^{p+1}H^{N-2}r^{N-1}\alpha'}{\sqrt{1-K}}\right]'-\frac{(N-1)B^{p+1}H^{N-2}r^{N-3}\sin 2\alpha}{2\sqrt{1-K}}=0.
\end{equation}
Armed with the energy-momentum tensor
\begin{eqnarray}
T_{AB}&=&\frac{2}{\sqrt{g}}\frac{\delta S}{\delta g^{AB}}\nonumber\\
          &=&\frac{2}{\sqrt{g}}\left(\frac{-1}{2}\sqrt{-g}g_{AB}{\cal{L}}+\sqrt{-g}\frac{\delta\cal{L}}{\delta g^{AB}}\right)\nonumber\\
          &=&-g_{AB}{\cal{L}}
          +\frac{\partial_{A}\phi_{i}\partial_{B}\phi_{i}}{\Bigg(1+\frac{1}{\beta^{2}_{N}}\partial_{C}\phi_{i}\partial^{C}\phi_{i}\Bigg)^{1/2}},
\end{eqnarray}
which yields
\begin{eqnarray}
T^{0}_{0}&=&-{\cal{L}},\nonumber\\
T^{r}_{r}&=&-{\cal{L}}-\frac{\alpha'^{2}}{H^{2}\sqrt{1-K}},\nonumber\\
T^{\theta}_{\theta}&=&-{\cal{L}}-\frac{\sin^{2}\alpha}{H^{2}r^{2}\sqrt{1-K}},
\end{eqnarray}
we are ready to solve the coupled Einstein's equations,
\begin{equation}
G^{\mu}_{\nu}=\kappa^{2}T^{\mu}_{\nu},
\end{equation}
with $\kappa^{2}\equiv 16\pi G_{D}$, along with (\ref{eq:chiralgrav}). 

 \subsection{Self-gravitating Solitonic DBI $p$-Branes}
 
\begin{figure}[htbp]
\centering\leavevmode
\epsfysize=8.5cm \epsfbox{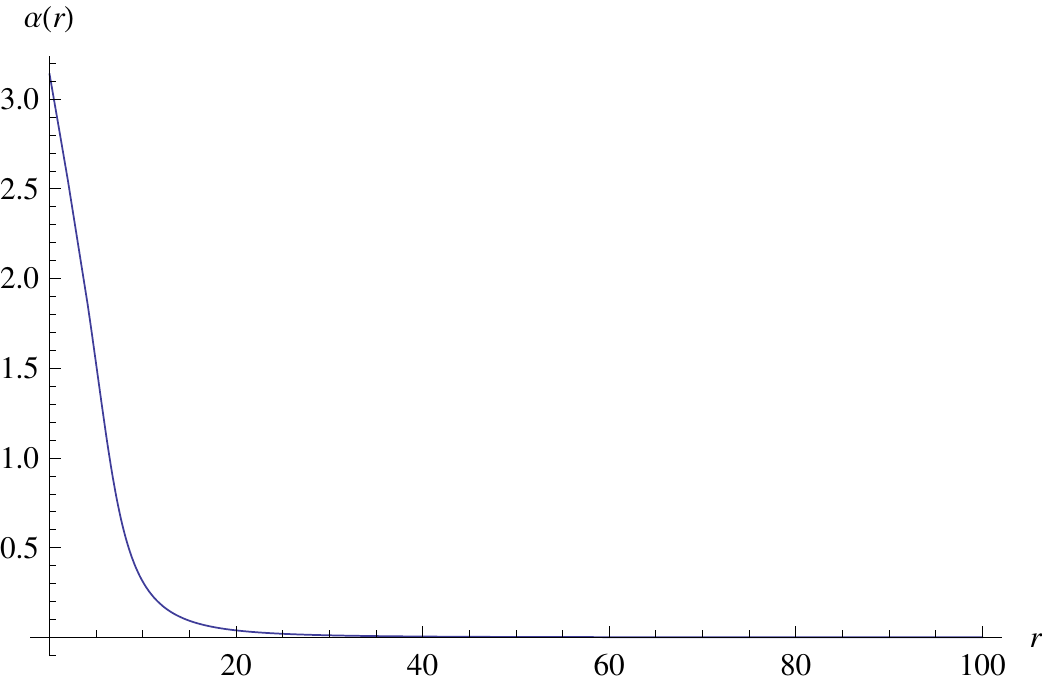}
\caption {Solitonic DBI $3$-brane in $(7+1)$-dimensions ($N=4$) with $\beta=1/2$.}
\label{N4}
\end{figure}

\begin{figure}[htbp]
\centering\leavevmode
\epsfysize=8cm \epsfbox{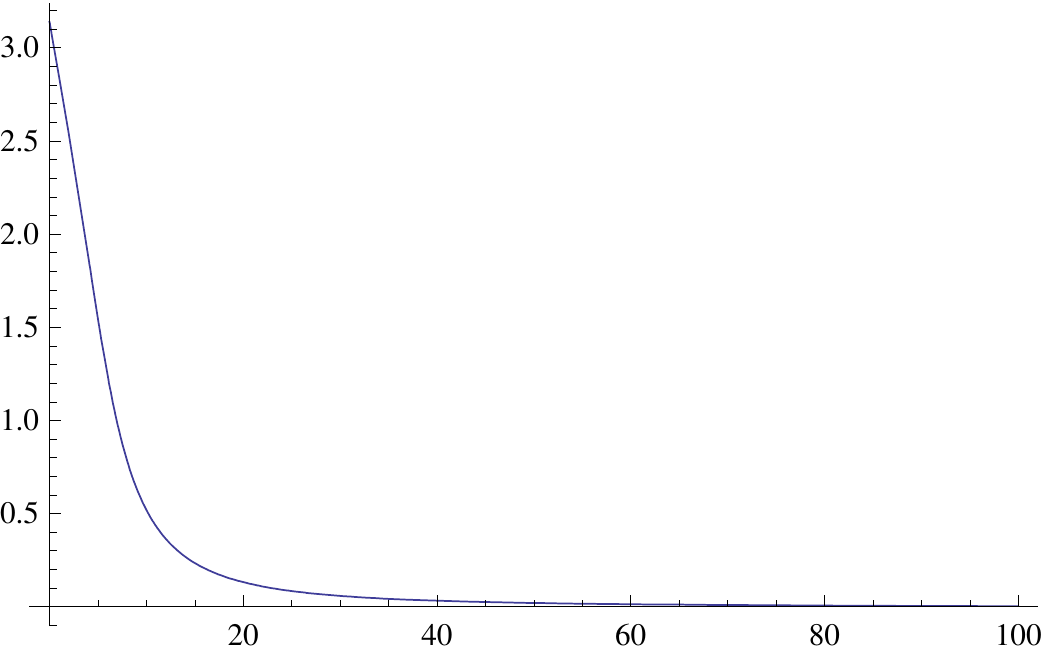}
\caption {Solitonic DBI $3$-brane in $(10+1)$-dimensions ($N=7$) with $\beta=1/2$.}
\label{N7}
\end{figure}

We solve the field equations numerically and show some of the solutions in figs.(\ref{N4})-(\ref{N7}). To be solitons the solutions must not be singular at the origin. Therefore we require regularity around the cores,
\begin{eqnarray}
\label{eq:regular}
\alpha(r)&\approx&\pi+\alpha_{1}\ r+O(r^{3}),\nonumber\\
B(r)&\approx&B_{0}+B_{2}\ r^{2}+O(r^{4}),\nonumber\\
H(r)&\approx&H_{0}+H_{2}\ r^{2}+O(r^{4}),
\end{eqnarray}
where $\alpha_{1}, B_{0}$, and $H_{0}$ are three undetermined constants\footnote{They become the shooting parameters in our numerical method.} in terms of which the higher-order coefficients of expansion can be expressed,
\begin{eqnarray}
B_{2}&=&-\frac{\kappa B_{0}\bigg(N\alpha_{1}^{2}-2\beta^{2}H_{0}^{2}+2\beta^{2}\sqrt{1-\frac{N\alpha_{1}^{2}}{\beta^{2}H_{0}^{2}}}H_{0}^{2}\bigg)}{2N(N+p-1)\sqrt{1-\frac{N\alpha_{1}^{2}}{\beta^{2}H_{0}^{2}}}},\nonumber\\
H_{2}&=&\frac{H_{0}\bigg(-2N\ p B_{2}-B_{0}H_{0}^{2}\beta^{2}\kappa+B_{0}H_{0}^{2}\beta^{2}\kappa\sqrt{1-\frac{N\alpha_{1}^{2}}{\beta^{2}H_{0}^{2}}}\bigg)}{2N(N-1)B_{0}}.
\end{eqnarray}
Notice that regularity requires the expansion of the chiral angle $\alpha(r)$ to consist only of odd-power terms and the metric functions, $B(r)$ and $H(r)$, to consist only of even-power terms.

\begin{figure}[htbp]
\centering\leavevmode
\epsfysize=8cm \epsfbox{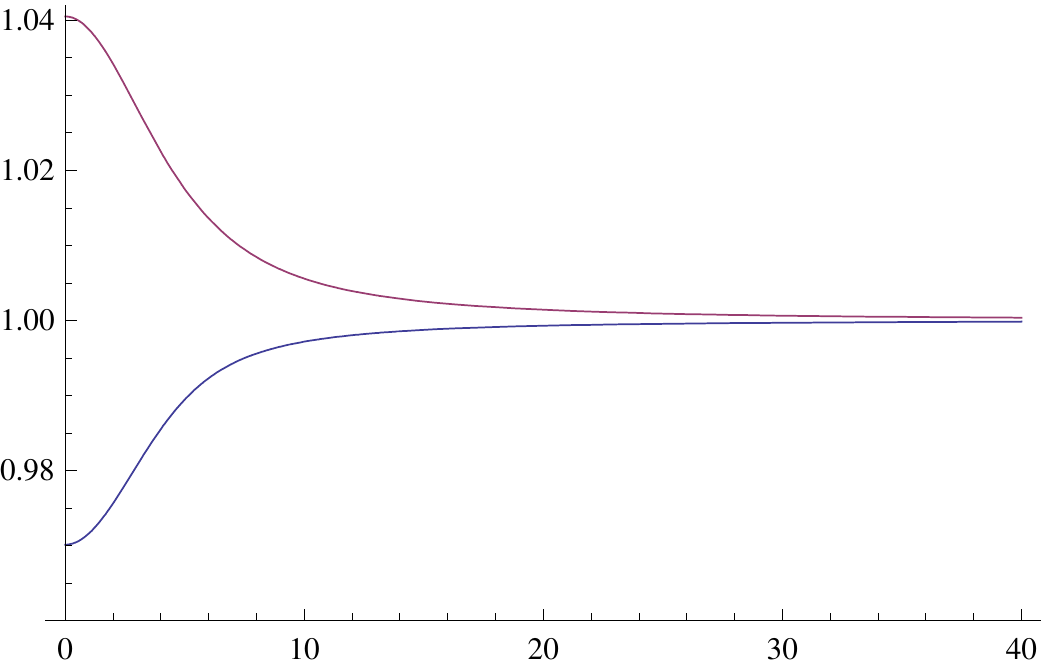}
\caption {Profiles of metric functions $B(r)$ (lower) and $H(r)$ (upper) in $(10+1)$-dimensions ($N=7$) with $\beta=1/2$. Both asymptote to the flat space.}
\label{ahmetric}
\end{figure}

An example of metric profiles is shown in fig.(\ref{ahmetric}). Here we have more input parameters {\it i.e.}, the number of dimensions on the brane ($p$) and perpendicular to it ($N$), and free parameters, {\it i.e.}, the gravitational coupling $\kappa$ and chiral coupling $\beta$. In this work we fix $p=3$ since we are only interested in the most realistic case, the case where we live on a $3$-brane. For each number of co-dimension $N$ we have two free parameters, $\kappa$ and $\beta$.

For each $N$-codimension, we investigate the dependence of solutions on $\beta$. The behavior resembles its flat space-time counterparts. The coupling constant $\beta$ controls the thickness of defects. Here we only show solutions with $\beta_{N}=1/2$ (strongly-coupled case). While their width is determined by $\beta$, their existence depends on $\kappa$. Variation of $\kappa$ does not change the width, but (as in the case of the Skyrme branes~\cite{ramadhan}) there exists a critical value, $\kappa_{crit}$, beyond which no solutions exist. The value of $\kappa_{crit}$ in each co-dimension $N$ depends also on $\beta$. For co-dimension $N=4$ and $\beta=1/2$ we found that $\kappa_{crit}\sim1/10$. Different $\beta$ gives different $\kappa_{crit}$.

Unlike the case  of Skyrme Branes~\cite{ramadhan} here we do not find any second branch of solutions upon coupling to gravity. We conjecture that this is caused by the nature of DBI form; {\it i.e.}, there is an upper bound value of $K$~(\ref{eq:K}) above which the Lagrangian becomes imaginary, thus non-physical. The {\it square-root} form of the Lagrangian suppresses the existence of upper branch solutions.

\begin{figure}[htbp]
\centering\leavevmode
\epsfysize=7.5cm \epsfbox{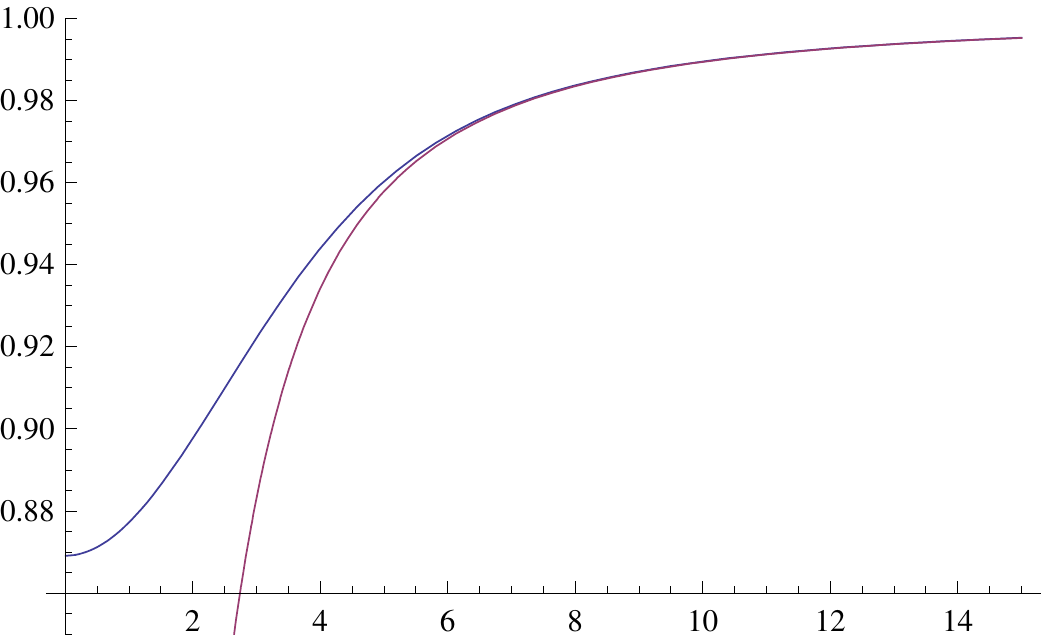}
\caption {Comparison between numerical and thin wall solutions (\ref{eq:bform}) of $B(r)$ for 3-branes (with $\beta=1/2$) in $(10+1)$-dimensions ($N=7$).}
\label{amatch}
\end{figure}

\begin{figure}[htbp]
\centering\leavevmode
\epsfysize=7.5cm \epsfbox{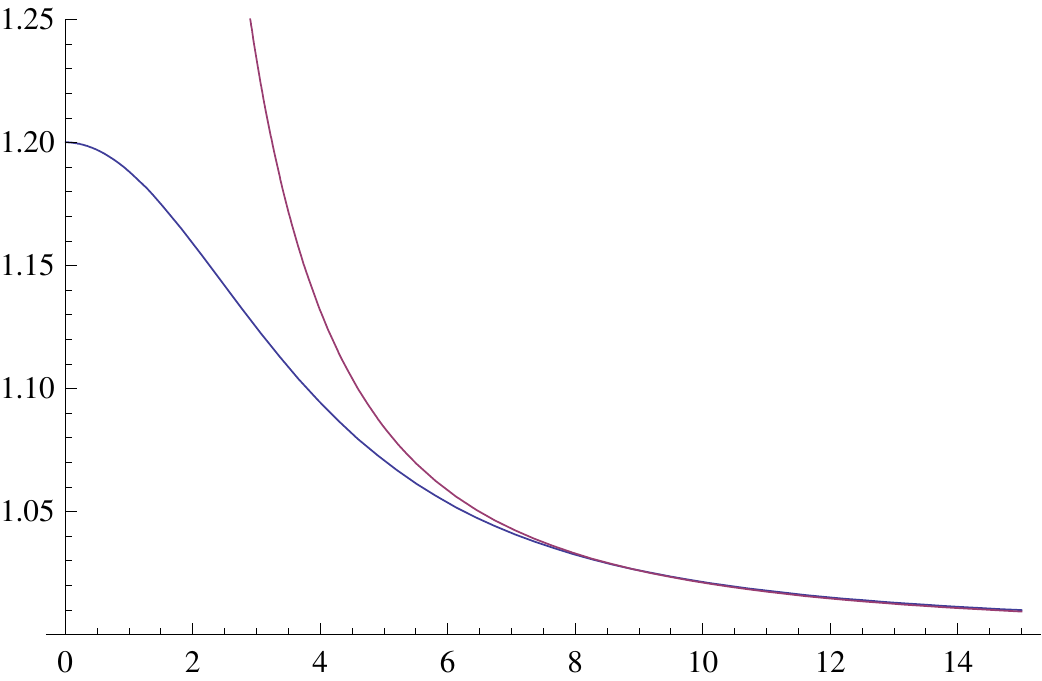}
\caption {Comparison between numerical and thin wall solutions (\ref{eq:hform}) of $H(r)$ for 3-branes (with $\beta=1/2$) in $(10+1)$-dimensions ($N=7$).}
\label{hmatch}
\end{figure}

The metric solutions, fig.(\ref{ahmetric}), are indeed the smooth and regular version of (\ref{eq:bform}) and (\ref{eq:hform}). To prove this, we employ the same technique as in~\cite{ramadhan} (see also~\cite{myers, gregoryemparan}); that is, we calculate the ADM mass~\cite{adm}, 
\begin{equation}
T^{ADM}_{\mu\nu}=\lim_{r\rightarrow\infty}\frac{1}{2\kappa^{2}}\oint d\Omega_{(N-1)}r^{(N-1)}\widehat{r}^{i}\left[\eta_{\mu\nu}\left(\partial_{i}h^{\sigma}_{\sigma}+\partial_{i}h^{j}_{j}-\partial_{j}h^{j}_{i}\right)
-\partial_{i}h_{\mu\nu}\right],
\end{equation}
which, for our isotropic gauge, gives
\begin{equation}
T^{ADM}_{\mu\nu}=\frac{\Omega_{(N-1)}}{2\kappa^{2}}\lim_{r\rightarrow\infty}r^{(N-1)}\left[pB(r)B'(r)+(N-1)H(r)H'(r)\right]\eta_{\mu\nu}.
\end{equation}
We can identify the single parameter $r_{0}$ that characterizes the p-brane solutions with its ADM mass for each given value of codimension. Plotting the vacuum and chiral-DBI branes solutions, figs.\ref{amatch}-\ref{hmatch}, they asymptotically match at large radius. Note that while the vacuum solutions are singular at finite distance from the origin ($r_{0}$), the numerical ones are regular everywhere. They start to deviate at $r\sim r_{0}$.
\section{Conclusions}

In this paper we have studied {\it extended} solitons in arbitrary $(N+1)$-dimensional non-linear sigma model. This is enabled by considering DBI-kinetic term in the Lagrangian. We showed that the solutions are stable under spatial rescaling and constructed, using numerical techniques, the explicit solutions having spherical symmetry ({\it i.e.}, hedgehog ansatz). They are characterized by the coupling constant $\beta$. For a fixed co-dimension $N$, the theory is weakly-coupled when $\beta>1$ and, upon Taylor expansion, can be reduced to the ordinary non-linear sigma model which have static (unstable) defects solutions, {\it textures}, for $\beta\rightarrow\infty$. On the other hand, for $0<\beta\leq 1$ the theory is now strongly-coupled and the non-linear terms becomes significant. We showed some solutions for several numbers of $N$ and values of $\beta$, and in principle any arbitrary $D$-dimensional solutions can be found. 

One motivation to study this theory is due to a conjecture put down by Gregory~\cite{gregory} that the naked singularity suffered by static boost-symmetric (uncharged) $D$-dimensional $p$-branes can be regularized by a judicious choice of core. While topological defects (solitons) seem to be the natural candidate, the asymptotically-flat condition severely restricts the option. It is well-known that neither domain wall~\cite{vilenkinwall} nor (global) string~\cite{gregstring} can gravitationally be static. The metric of a global monopole, on the other hand, is static but has a deficit solid angle and thus cannot be asymptotically-flat~\cite{barriola}. Textures become an interesting alternative, but unless some mechanism is introduced to stabilize it under spatial rescaling, the status of its static gravitational field is unclear. In~\cite{ramadhan} it was shown that by considering the Skyrmions, as the solitonic textures with non-canonical kinetic terms, we numerically proved the conjecture for $p=3$ and $N=3$. One purpose of this paper is to present the completion of the proof for arbitrary $N$ in a particular (string-theory-inspired) model.

The existence of the self-gravitating solitonic $p$-branes for each dimension depends on the value of $\kappa$, in a sense that there exists $\kappa_{crit}$ beyond which no static solutions found. We conjecture that the fate of super-critical configuration is either (i) collapsing to DBI black branes, or (ii) non-static ({\it i.e.}, inflating) DBI branes. At the moment we have yet investigated and hope to explore these possibilities in the future. 

The asymptotic flatness of the solutions provide an interesting possibility of realizing the Dvali-Gabadadze-Porrati (DGP)~\cite{dgp} braneworld model in higher dimensions. There has recently been some studies of DGP-realization in models with bulk topological defects fields (see~\cite{RR, FR, kolanovic1, kolanovic2}). Indeed in ref.~\cite{ringevaldim} it was shown that a braneworld model with (hyper-)monopole core in arbitrary dimensions can support the existence of metastable massive graviton which is essential in the realization of DGP theory. This proposal shed a light on the development of the higher-dimensional DGP models. Finding a metastable graviton in our theory can be developed along that way.  
\acknowledgments

We thank Jose Blanco-Pillado for bringing this topic into our attention and proof-read the manuscript. We also thank Noriko Shiiki and Eray Sabancilar for enlightening discussions. This draft was completed while I was a visitor at KEK theory group, Japan. I thank the institution for their hospitality.


\end{document}